\documentclass[conference]{IEEEtran}
\IEEEoverridecommandlockouts

\usepackage{cite}
\usepackage{amsmath,amssymb,amsfonts}
\usepackage{algorithm}
\usepackage{algpseudocode}
\usepackage{graphicx}
\usepackage{textcomp}
\usepackage{xcolor}
\def\BibTeX{{\rm B\kern-.05em{\sc i\kern-.025em b}\kern-.08em
    T\kern-.1667em\lower.7ex\hbox{E}\kern-.125emX}}

\usepackage{amsmath,amssymb,amsfonts}

\usepackage{graphicx, color, caption, subcaption}
\graphicspath{{fig/}}
\usepackage{mathtools}

\DeclarePairedDelimiterX{\infdivx}[2]{(}{)}{%
	#1\;\delimsize\|\;#2%
}

\usepackage{pgf}
\usepackage{bm}

\newcommand{\guillemets}[1]{``#1''}
\newcommand{\set}[1]{\left\{#1\right\}}

\newcommand{\myvector}[1]{\bm{#1}}

\newcommand{\mymatrix}[1]{\bm{#1}}
\newcommand{\entry}[1]{\left[{#1}\right]}
\newcommand{\hermitian}{H}

\newcommand{\R}{\mathbb{R}} % Ensemble R
\newcommand{\C}{\mathbb{C}} % Ensemble C
\newcommand{\E}{\mathrm{E}} % Espérance

\newcommand{\eqdef}{\triangleq}

\newcommand{\norm}[1]{\left\lVert#1\right\rVert}

\newcommand{\ie}{{i.e.},}

\DeclareMathOperator{\diag}{\mathrm{diag}}
\DeclareMathOperator*{\vect}{\mathrm{vec}}

\DeclareUnicodeCharacter{2212}{-}

\begin{document}

\title{Moment-based Characterization of Spatially Distributed Sources in SAR Tomography
\thanks{This work was partially supported by DGA/AID project 2022 65 0082.}
}

\author{Colin Cros\textsuperscript{1}, Laurent Ferro-Famil\textsuperscript{1,2}\\
\textsuperscript{1} ISAE Supaero, University of Toulouse, France \\
\textsuperscript{2} CESBIO, University of Toulouse, France \\
\texttt{firstname.lastname@isae-supaero.fr}
}

\maketitle\texttt{}

\begin{abstract}
This paper presents a non-parametric method for 3-D imaging of natural volumes using Synthetic Aperture Radar tomography. This array processing-based technique aims at characterizing a spatially distributed density of incoherent sources, whose shape is imprecisely known. The proposed technique estimates the moments of the reflectivity density using a low-complexity covariance matching approach, and retrieves the mean location, dispersion, and power of the distributed source. Numerical simulations of realistic tomographic scenarios show that the proposed model-free scheme achieves better accuracy than slightly misspecified maximum likelihood estimators, derived from approximately known distribution shapes.
\end{abstract}


\section{Introduction}

Synthetic Aperture Radar (SAR) tomography represents a unique tool for
characterizing natural environments at a large scale from their 3-D
electromagnetic reflectivity. Its is particularly well adapted to the
monitoring of forests \cite{aghababaei2020forest} and is an operating mode of ESA's
upcoming BIOMASS mission, based on a spaceborne P-band ($\lambda
\approx 70$cm) radar.  
% The  deployment of new satellite sensors for SAR imaging opens up new
% perspectives in Earth observation. In particular, the increased number
% of passes allows the construction of three-dimensional tomographic
% images of forests \cite{aghababaei2020forest} for better monitoring of
% their biomass. This technique provides vertical profiles of forest
% reflectivity by treating forests as diffuse sources of incoherent
% scatterers.
Array processing methods may be used to estimate the
elevation and reflectivity of discrete sources, using a small number of
irregularly spaced 2-D SAR acquisitions \cite{Gini_Lombardini_2005}. The characterization
of natural environments having a continuous density
of reflectivity, such as forests, may be severely affected by the
limited resolution performance associated with short arrays.  As suggested
in \cite{bou2024tropical} for a tropical forest observed at P-band,
model-based approaches using moderately spread sources may be used to estimate the elevation of the tree canopy, but cannot reliably determine the actual shape of the
reflectivity density.  This paper addresses the problem of characterizing a narrow
diffuse source, whose exact distribution is unknown, using a
potentially irregular antenna array. The goal is to determine the mean direction,
dispersion, and power of the source, i.e., the height, and thickness and
reflectivity of the forest canopy, respectively.

Characterizing a diffuse source using array processing is a
recurring problem in the literature. For known distribution shapes,
efficient alternatives to the Maximum Likelihood (ML) estimator
\cite{stoica1990performance} have been proposed using the COMET-EXIP estimator \cite{besson2000decoupled,
  zoubir2006modified}. Furthermore, the moments of the distribution
have been used to retrieve the arrival angle
\cite{shahbazpanahi2004covariance}, or the dispersion of the
sources. The latter is used, for example, in the generalizations of
the MUSIC and ESPRIT algorithms for diffuse sources
\cite{valaee1995parametric, shahbazpanahi2001distributed}. However, to
the best of our knowledge, the central moments have never been used to
characterize the entire distribution of a source with an unknown
distribution.\\
This paper proposes a new method to characterize a diffuse source
based on a COMET estimator and on the moments of the distribution,
without assuming any source model. It is shown that the proposed
approach can accurately reconstruct the source characteristics,
whereas incorrectly parameterized model-based methods give highly biased results, especially for the source power.

The remainder of this letter is organized as
follows. Section~\ref{sec: Problem} introduces the problem
statement. Section~\ref{sec: Proposed method} presents a new
estimation scheme based on the distribution moments. Section~\ref{sec:
  Simulation} describes the advantages of this new scheme in simulations. Finally, Section~\ref{sec: Conclusion} concludes this paper.

\section{Problem statement} \label{sec: Problem}
\subsection{Tomographic signal model}

\begin{figure}
	\centering
	\includegraphics{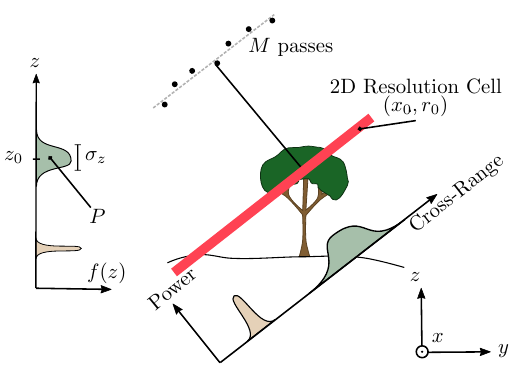}
	\caption{SAR Tomography principle.}
    \label{fig: Illustration SAR tomography}
\end{figure}

A tomographic SAR measurement, illustrated in Figure~\ref{fig: Illustration SAR tomography}, consists of $M$ 2D SAR acquisitions performed from slightly shifted, and ideally parallel, tracks. Under the widely accepted first order Born approximation, considering that the echoes of individual scatterers sum up to form the response a scene, the expression of one of the 2D SAR images of a general 3D environement may be written as:
\begin{equation}
  \label{eq:2D_SAR}
  s_m(x_0,r_0) = \int h(x-x_0,r-r_0) a_c(\mathbf{r})  \text{e}^{j
    k_{c} r} d\mathbf{r} + \epsilon_m(x_0,r_0), 
\end{equation}
where $r_0$ and $x_0$ stand for the range and azimuth coordinates of
the images, respectively, $h(x,r)$ is the 2D SAR ambiguity
function, $k_{c}$ is the carrier two-way wavenumber, $\myvector{r}$ is a 3D
coordinate, and $r$ is its associated range, \ie{}  the orthogonal
distance between the flight track and location $\myvector{r}$,
$\epsilon_m(x_0,r_0)$ is the normally distributed focused acquisition noise, and
$a_c(\myvector{r})$ represents the coherent density of scattering of
the scene. Over natural environments, this density, consisting of a
large number of independent contributions originating from randomly
distributed sources, is usually assumed to follow a complex circular
normal distribution verifying
$ \E[a_c(\mathbf{r})] = 0$,
$\E[a_c(\mathbf{r})a_c(\mathbf{r}+d\mathbf{r})] = 0 $, and $
\E[a_c(\mathbf{r})a_c^*(\mathbf{r}+d\mathbf{r})]
=\sigma_a^2(\mathbf{r})\delta(d\mathbf{r})$ 
% \begin{subequations}
% 	\begin{align}
% 		\E[a_c(\mathbf{r})] &= 0, \\
% 		\E[a_c(\mathbf{r})a_c(\mathbf{r}+d\mathbf{r})] &= 0, \\
% 		\E[a_c(\mathbf{r})a_c^*(\mathbf{r}+d\mathbf{r})] &=\sigma_a^2(\mathbf{r})\delta(d\mathbf{r}),
% 	\end{align}
% \end{subequations}
with $\sigma_a^2(\mathbf{r})$, the 3D density of reflectivity of the scene.

The ambiguity associated with 2D SAR imaging may be perceived in \eqref{eq:2D_SAR}, with a cylindrical projection the 3D scattering density onto a 2D range-azimuth domain. SAR tomography proposes to overcome this limitation using spatial diversity and coherent array processing. To do so, the $M$ SAR images are coregistered to a reference geometric frame, and are demodulated to the spatial baseband domain. The resulting tomographic image stack, $\mathbf{y}(x_0,r_0) \in \C^M$ follows a centered complex circular distribution, $\mathbf{y} \sim \mathcal{CN}(\mathbf{0},\mathbf{R})$, whose second order statistics can be used to retrieved 3D reflectivity features.  In the case of a horizontally homogeneous environment, i.e., when the 2D ambiguity function is narrow enough so that the scattering density at a given elevation may be considered as stationary within a 2D resolution cell, the expression of the interferometric cross-correlation simplifies to:
\begin{equation}
  \label{eq:y_xcorr}
  \E[y_n y_m^*] = \int f(z) \text{e}^{j k_{z_{nm}} z}  dz + \sigma_\varepsilon^2 \delta(n-m),
\end{equation}
where the range and azimuth coordinates of the concerned 2D resolution cell have been omitted for the sake of clarity, and $f(z)$, the vertical component of $\sigma_a^2(\mathbf{r})$, represents the vertical density of reflectivity. The relationship between the elevation of a contributing source and the interferometric phase of its response is given by $ k_{z_{nm}} = k_{z_{n}}-k_{z_{m}}$, with  $k_{z_{n}}$ the interferometric wave number of the n\textsuperscript{th} image. The vertical Fourier resolution is given by $\delta z = 2 \pi/(\max_n k_{z_{n}} - \min_n k_{z_{n}})$, whereas the vertical ambiguity may be approximated, for a quasi-uniform wave number spacing, as $z_{amb} \approx M \delta z$.

\subsection{Covariance matrix model}

The tomographic cross-correlation expression of \eqref{eq:y_xcorr} may be generalized using antenna array processing formalism. The covariance matrix of the tomographic acquisition can be formulated as:
\begin{equation}\label{eq: Definition R}
	\mymatrix{R} \eqdef \E[\myvector{y} \myvector{y}^{\hermitian}]
        = P \int p(\omega - \omega_0) \mymatrix{a}(\omega)
        \mymatrix{a}(\omega)^{\hermitian} \, d \omega + \sigma_{\varepsilon}^2 \mymatrix{I}_M.
\end{equation}
where $\omega_0$ denotes the mean spatial frequency, $P$ the power of the source, $p$ its distribution around $\omega_0$ which satisfies $\int p(\omega)\, d\omega = 1$ and $\int \omega p(\omega)\, d\omega = 0$, and $\mymatrix{a}(\omega)$ is a steering vector expressed as:
\begin{equation}\label{eq: Steering vector}
	\myvector{a}(\omega) \eqdef \begin{pmatrix}1 & e^{j 2\pi u_2 \omega} & \cdots & e^{j 2\pi u_M \omega}\end{pmatrix}^{\intercal} \in \C^M.
\end{equation}
In \eqref{eq: Steering vector}, $u_k$ denotes the distance (in wavelength) between the first and the $k$\textsuperscript{th} element in the equivalent linear array, and $\omega = \sin \phi$, with $\phi$ the associated angle of arrival on the array.
The expression of $\mymatrix{R}$, \eqref{eq: Definition R}, may be reformulated as:
\begin{subequations}\label{eq: Covariance matrix}
\begin{equation}
\mymatrix{R} =  \mymatrix{a}(\omega_0) \mymatrix{a}^{\hermitian}(\omega_0) \odot P \mymatrix{B} + \sigma_{\epsilon}^2 \mymatrix{I}_M,  
\end{equation}
where $\odot$ denotes the Hadamard (element-wise) matrix pro\-duct, and $\mymatrix{B}$ is a form matrix that depends on the scattering distribution. Its $(k,l)$ entry is:
\begin{equation}
	\entry{\mymatrix{B}}_{k,l} \eqdef \int p(\tilde \omega) e^{j 2\pi (u_k-u_l) \tilde \omega} \, d\tilde\omega,
\end{equation}
\end{subequations}
where %the deviation
$\tilde\omega \eqdef \omega - \omega_0$.
%has been introduced to lighten the expression.
The objective of the analysis is to estimate, from the covariance
matrix of the observations $\set{\myvector{y}}$, three characteristics
of the power density of the distributed source: $(i)$ its mean spatial
frequency $\omega_0$, $(ii)$ its standard deviation $\sigma_\omega$,
and $(iii)$ its power $P$. In the context of forest SAR tomography,
they correspond to the height $z_0$ of the forest canopy, its thickness $\sigma_z$, and its reflectivity $P$ as illustrated in Figure~\ref{fig: Illustration SAR tomography}.

\section{Proposed estimation scheme}\label{sec: Proposed method}

\subsection{COMET estimator}

%Since there is no information contained in the first order moments of the measurements $\myvector{y}(t)$, the estimation scheme relies on the estimation of the covariance matrix of the measurements \eqref{eq: Covariance matrix}.
The COMET estimator \cite{ottersten1998covariance} is used to estimate
the characteristics of the source. It is based on a generalized
least squares method, and represents an efficient alternative to the
Maximum Likelihood estimator of the unconditional model at hand \cite{stoica1990performance}. If the covariance matrix is parameterized by some vector $\myvector{\theta}$ as $\mymatrix{\hat R}(\myvector{\theta})$, the COMET estimate is obtained by minimizing the following cost function:
\begin{equation}\label{eq: Cost function}
	J(\myvector{\theta}) = \norm{\mymatrix{W}^{\hermitian/2} (\mymatrix{\bar R} - \mymatrix{\hat R}(\myvector{\theta}))\mymatrix{W}^{1/2}}^2,
\end{equation}
where $\mymatrix{\bar R} \eqdef \frac{1}{N}\sum_{t=1}^N \myvector{y}(t)\myvector{y}(t)^{\hermitian}$ is the sample covariance matrix, $\myvector{y}(t)$ is a realization of $\myvector{y}$, and $\mymatrix{W}$ is a Hermitian weight matrix. Two choices of weight matrix are commonly used: $\mymatrix{W} = \mymatrix{I}$ which corresponds to an unweighted least square method, and $\mymatrix{W} = \mymatrix{\bar {R}}^{-1}$ which is proved to induce an asymptotically efficient estimator if the model $\mymatrix{\hat R}(\myvector{\theta})$ is correctly specified \cite{ottersten1998covariance}.
Before applying the COMET estimator, a model should be chosen for the estimated matrix $\mymatrix{\hat R}(\myvector{\theta})$.

\subsection{Moment-based model}

The difficulty in modeling the covariance matrix $\mymatrix{R}$ comes
from the fact that the scatterer distribution $p$ is unknown. The
entries of the matrices $\mymatrix{R}$ and $\mymatrix{B}$ are not
directly related to the distribution $p$, but to its Fourier Transform
(FT) $\widehat{p}$, described here using a non-parametric model.
Assuming that the scatterer distribution is narrow, the FT
spreads over a large domain, and is very smooth at $0$. Therefore,
${\widehat{p}}$ can be accurately approximated near the
origin by the first terms of its Taylor expansion, and represents, by definition,
the characteristic function of the distribution. These coefficients correspond, up to a factor, to the central moments
of the distribution. Let $\mu_d$ denote the $d$-th central moment of
the distribution, $\mu_d \eqdef \E[\tilde \omega^d]$, then $\mu_0 = 1$
and $\mu_1 = 0$.
It is proposed to parameterize $\widehat{p}$
by its first $D$ central moments, the choice of the order $D$ being 
discussed in Section~\ref{sec: Simulation}. For a vector
$\myvector{\mu} = \begin{pmatrix}\mu_2 & \cdots &\mu_D\end{pmatrix}^{\intercal}
\in \R^{D-1}$, the FT $\xi \mapsto \widehat{p}(\xi)$ is approximated 
by: 
\begin{equation}\label{eq: Model TF p}
	\widehat{p}(\xi, \myvector{\mu}) = 1 + \sum_{d=2}^D \frac{j^d}{d!} \mu_d \xi^d.
\end{equation}
Such a parameterization has two main advantages. First, it does not assume any scatterer distribution model, which would lead to a loss of accuracy in the case of an incorrect specification, as illustrated in Section~\ref{sec: Simulation}. Second, the central moment $\mu_2 \eqdef \sigma_\omega^2$ corresponds to the dispersion of the distribution to be estimated.

The parametrization \eqref{eq: Model TF p} induces the following model for the matrices $\mymatrix{B}$ and $\mymatrix{R}$:
\begin{subequations}
	\begin{align}
		\mymatrix{\hat{B}}(\myvector{\mu}) &= \myvector{1}\mymatrix{1}^{\intercal} +  \sum_{d=2}^D \frac{j^d}{d!} \mu_d \mymatrix{U}^{(d)}, \\
		\mymatrix{\hat{R}}(\myvector{\theta}) &= \myvector{a}(\omega_0)\myvector{a}(\omega_0)^{\hermitian} \odot P \mymatrix{\hat{B}}(\myvector{\mu}) + \sigma_{\varepsilon}^2 \mymatrix{I}_M,
	\end{align}
\end{subequations}
where the parameter vector is $\myvector{\theta} \eqdef \left(\omega_0,\, P,\, \sigma_{\varepsilon}^2,\, \myvector{\mu}^{\intercal} \right)^{\intercal} \in \R^{D+2}$, and the matrices $\mymatrix{U}^{(d)}$ depend only on the geometry of the array. Its $(k,l)$ entry is $\entry{\mymatrix{U}^{(d)}}_{k,l} = (u_k - u_l)^d$.
In particular, this parametrization is also valid for non-uniform arrays.

\subsection{Estimation algorithm}

Before applying the COMET estimation scheme, a transparent
modification is made in the parameter vector to ease the
optimization. The vector $\myvector{\nu} \eqdef P \myvector{\mu}$ is
used to define the parameter vector $\myvector{\theta} \eqdef \left(\omega_0,\, P,\, \sigma_{\varepsilon}^2,\, \myvector{\nu}^{\intercal} \right)^{\intercal} \in \R^{D+2}$. This change of variable makes the dependence of $\mymatrix{\hat R}$ linear on all the parameters, except $\omega_0$. Note that as $P > 0$, this change of variable is invertible.

Denote, as in \cite{ottersten1998covariance}, $\myvector{\alpha} \eqdef \left(P, \sigma_{\varepsilon}^2, \myvector{\nu}^{\intercal}\right)^{\intercal} \in \R^{D+1}$ the set of linear parameters in $\myvector{\theta}$. The cost function $J$, \eqref{eq: Cost function}, can be re-expressed in vector form as:
\begin{equation}\label{eq: Cost vector form}
	J(\omega_0, \myvector{\alpha}) = (\myvector{\bar r} - \mymatrix{\Psi}(\omega_0) \mymatrix{J} \myvector{\alpha})^{\hermitian} \left(\mymatrix{W}^{\intercal} \otimes \mymatrix{W}\right) (\myvector{\bar r} - \mymatrix{\Psi}(\omega) \mymatrix{J} \myvector{\alpha}),
\end{equation}
where $\myvector{\bar r} \eqdef \vect \mymatrix{\bar{R}}$ is the vector obtained by stacking the columns of $\mymatrix{\bar{R}}$, $\otimes$ denotes the Kronecker matrix product, $\mymatrix{J}$ is the matrix such that $\vect\left\{P\mymatrix{B}(\mu) + \sigma_{\varepsilon}^2 \mymatrix{I}_M\right\} = \mymatrix{J} \myvector{\alpha}$, and:
\begin{align*}
	\mymatrix{\Phi}(\omega) &\eqdef \diag \myvector{a}(\omega), & \mymatrix{\Psi}(\omega) &\eqdef \mymatrix{\Phi}(\omega)^{\hermitian} \otimes \mymatrix{\Phi}(\omega).
\end{align*}
As pointed out in \cite{ottersten1998covariance}, for any set $\omega_0$ value, \eqref{eq: Definition R} is a quadratic form in $\myvector{\alpha}$ whose minimum is achieved at:
\begin{multline}\label{eq: Solution alpha}
	\myvector{\hat \alpha}(\omega_0) = \left[(\mymatrix{\Psi}(\omega_0) \mymatrix{J})^{\hermitian} (\mymatrix{W}^{\intercal}\otimes\mymatrix{W})(\mymatrix{\Psi}(\omega_0)\mymatrix{J})\right]^{-1}\\
	\times \mymatrix{\Psi}(\omega_0) \mymatrix{J})^{\hermitian} (\mymatrix{W}^{\intercal}\otimes\mymatrix{W})\myvector{\bar r}.
\end{multline}
Thus, a single optimization on $\omega_0$ is required to find the minimizer of $J$. By re-injecting \eqref{eq: Solution alpha} in $J$, one obtains:
\begin{equation}\label{eq: Solution omega}
	\hat \omega_0 = \arg \max \myvector{y}(\omega)^{\hermitian} \mymatrix{Y}(\omega)^{-1} \myvector{y}(\omega),
\end{equation}
with:
\begin{align*}
	\myvector{y}(\omega) &\eqdef (\mymatrix{\Psi}(\omega) \mymatrix{J})^{\hermitian} (\mymatrix{W}^{\intercal}\otimes\mymatrix{W})\myvector{\bar r}, \\
	\mymatrix{Y}(\omega) &\eqdef (\mymatrix{\Psi}(\omega) \mymatrix{J})^{\hermitian} (\mymatrix{W}^{\intercal}\otimes\mymatrix{W})\mymatrix{\Psi}(\omega)\mymatrix{J}.
\end{align*}
The one-dimensional optimization problem in \eqref{eq: Solution omega}
can be solved with a linear search or a golden-section search. Once, $\hat\omega_0$ has been estimated, all the other parameters can be derived. The full estimation algorithm is summarized in Algorithm~\ref{alg: Estimation algorithm}.
\begin{algorithm}
	\caption{Estimation algorithm.}
	\begin{algorithmic}[1]
		\State Form the sample covariance matrix $\mymatrix{\bar{R}}$ from the measurements $\set{\myvector{y}}$.
		\State Estimate $\hat \omega_0$ by solving the one-dimensional optimization problem \eqref{eq: Solution omega}.
		\State Compute $\hat P$, $\hat \sigma_{\varepsilon}$, and $\myvector{\hat \nu}$ from \eqref{eq: Solution alpha} with $\omega_0 = \hat\omega_0$.
		\State Compute the estimates of the central moments as: $\hat \mu = \hat \nu / \hat P$. In particular, $\hat \sigma_\omega^2 = \hat \mu_2$.
	\end{algorithmic}\label{alg: Estimation algorithm}
\end{algorithm}

\section{Numerical simulations}\label{sec: Simulation}

This section presents numerical simulations to demonstrate the
efficiency of the proposed approach. The simulation scenario
illustrates a typical SAR tomography configuration. The observed scene
corresponds to a forest whose ground response has been canceled
out using a notching preprocessing step
\cite{Mariotti_notch_2020}. Unless otherwise stated the selected
configuration is characterized by its height ambiguity $z_{amb}=100$ m, its vertical resolution $\delta z \approx
14.3$ m, canopy thickness $\sigma_z$ of $5$ m, and number of acquisitions $M=7$, which are typical settings for forest SAR tomography \cite{bou2024tropical}. The weight matrix is fixed to $\mymatrix{W} = \mymatrix{\bar{R}}^{-1}$. 

\subsection{Influence of the order $D$}

\begin{figure*}
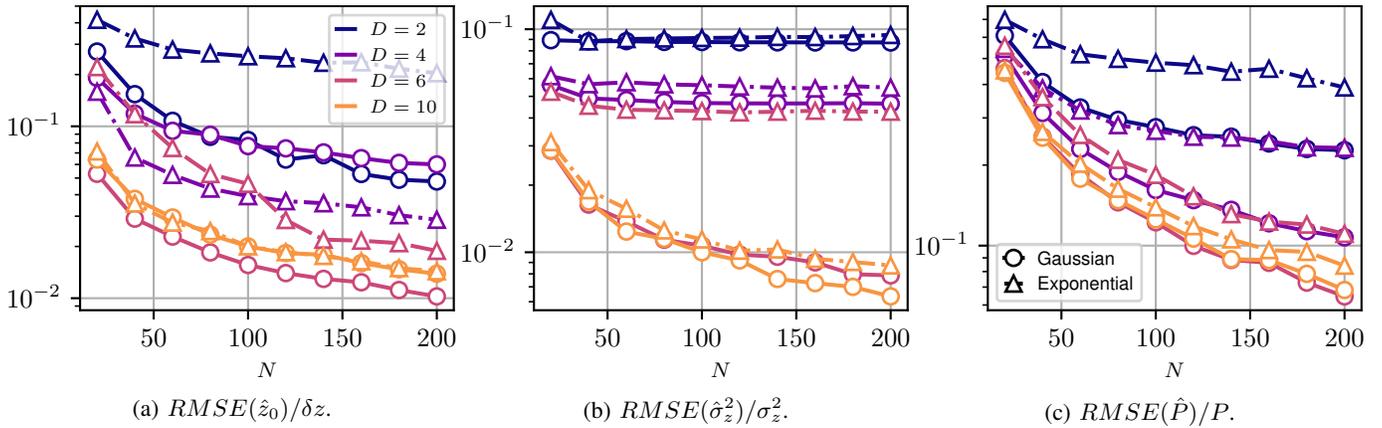

	\centering
	\null\hfill
	\subfloat[$RMSE(\hat z_0)/\delta z$.\label{sfig: RMSE on omega_0}]{\input{fig/comparison_order_omega_0.pgf}}
	\hfill
	\subfloat[$RMSE(\hat \sigma_z^2)/\sigma_z^2$.\label{sfig: RMSE on sigma_o}]{\input{fig/comparison_order_sigma_o.pgf}}
	\hfill
	\subfloat[$RMSE(\hat P)/P$.\label{sfig: RMSE on P}]{\input{fig/comparison_order_P.pgf}}
	\hfill\null
	\caption{RMSE for different orders $D$ and different (Gaussian
          and exponential) distribution shapes. The simulation
          parameters are $\sigma_z = 5$ m, $SNR =
          P/\sigma_{\varepsilon}^2 = 20$ dB,  $M = 7$ sensors
          ($D_\text{max} = 11$), $\delta z \approx 14.3$ m. The RMSE
          estimates were computed over $5\ 000$ realizations} 
	\label{fig: Comparison order}
\end{figure*}

%Let us start by discussing the choice of the maximum order $D$.
%First of all,
The possible values of $D$ are constrained by the size of the
array. With an $M$-sensor array, the covariance matrix
$\mymatrix{R}$ has up to $M(M-1) + 1$ different entries. This number
is reduced to only $2M-1$ in the case of a uniform array. The number
of parameters in $\myvector{\theta}$ must be smaller than the number
of observables to ensure the solvability of the problem. Therefore,
the maximum value of $D$ is: 
\begin{equation}
	D_{\mathrm{max}} = \left\{\begin{array}{cl}
		2M - 3 & \text{for a uniform array,}\\
		M(M-1) - 1 & \text{for a general array.}
	\end{array}\right.
\end{equation} 
Thus, the hyper-parameter ${D}$ must satisfy $2 \le D \le D_{\mathrm{max}}$. In particular, identifying the dispersion of the source $\sigma_\omega$, requires at least $M = 3$ sensors.

A natural question is then, should ${D}$ be set to its maximum possible value? 
Conceptually, the moment-based estimation scheme is a polynomial
interpolation of the characteristic function of the scatterer
distribution, and $D$ corresponds to the order of interpolation. 
On the one hand, large orders allow to capture more
complex functions. However, large orders also increase the risk of
overfitting the data and oscillations, due to Runge's phenomenon.
In the context of SAR tomography, the number of sensors $M$ is small,
typically $7$ at most, and these two phenomena hardly occur. Hence,
larger $D$ values provide better accuracy.
Figure~\ref{fig: Comparison order} shows the evolution of the Root
Mean Square Error (RMSE) on the estimates as a function of the number
of snapshots $N$ and of the order $D$. A clear gain of performance is
obtained as the order increases.
It is important to note that the maximum order achieves similar
accuracy with two very different, Gaussian and Exponential,
distribution shapes. This  confirms that the proposed approach is
independent of the true distribution, and in particular, that it works
similarly with symmetric and asymmetric distributions.

\subsection{Comparison with parametric schemes}

\begin{figure*}
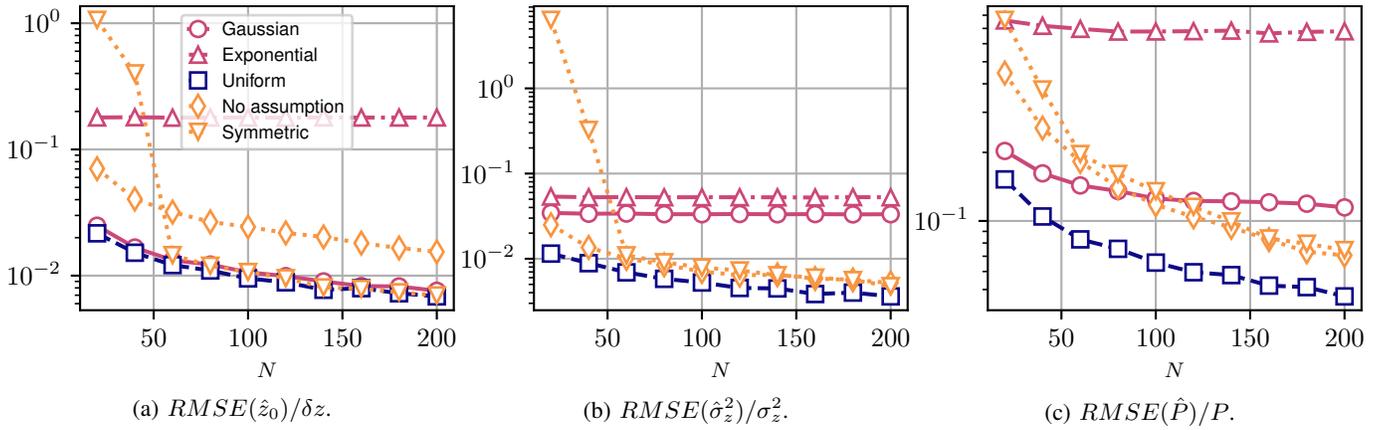

	\centering
	\null\hfill
	\subfloat[$RMSE(\hat z_0)/\delta z$.]{\input{fig/comparison_ML_omega_0.pgf}}
	\hfill
	\subfloat[$RMSE(\hat \sigma_z^2)/\sigma_z^2$.]{\input{fig/comparison_ML_sigma_o.pgf}}
	\hfill
	\subfloat[$RMSE(\hat P)/P$.]{\input{fig/comparison_ML_P.pgf}}
	\hfill\null
	\caption{RMSE on the estimation of a uniform distributed
          source for different estimation schemes: ML estimators under
          three different distribution assumptions
          (\guillemets{Gaussian}, \guillemets{Exponential} and
          \guillemets{Uniform}), the proposed approach with $D =
          D_{\text{max}}$ (\guillemets{No assumption}), and its
          adaptation to symmetric distributions
          (\guillemets{Symmetric}). The same configuration as in
          Fig.~\ref{fig: Comparison order} is used.}
	\label{fig: Comparison parametric methods}
\end{figure*}

\begin{figure*}
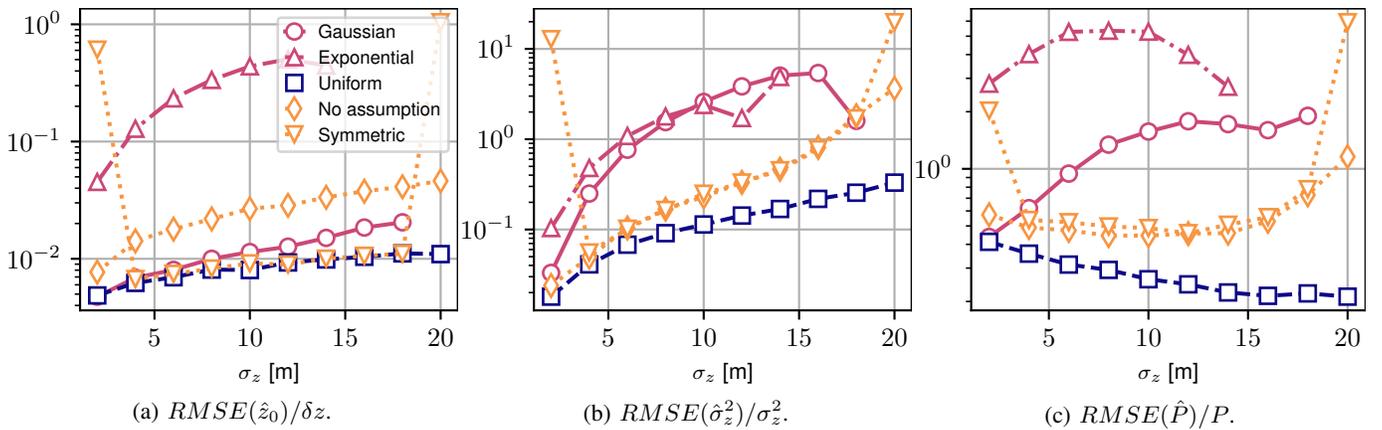

	\centering
	\null\hfill
	\subfloat[$RMSE(\hat z_0)/\delta z$.]{\input{fig/comparison_sigma_omega_0.pgf}}
	\hfill
	\subfloat[$RMSE(\hat \sigma_z^2)/\sigma_z^2$.]{\input{fig/comparison_sigma_sigma_o.pgf}}
	\hfill
	\subfloat[$RMSE(\hat P)/P$.]{\input{fig/comparison_sigma_P.pgf}}
	\hfill\null
	\caption{RMSE as a function of the source spread
          $\sigma_z$. The same configuration as in Fig.~\ref{fig:
            Comparison parametric methods} is used, the number of
          snapshots is set to $N=100$. For large $\sigma_z$ values,
          the misspecified ML algorithms did not converge.} 
	\label{fig: Comparison function of sigma}
\end{figure*}

The proposed approach is compared with parametric methods. Figure~\ref{fig:
  Comparison parametric methods} presents  RMSE values 
for different estimation schemes. The proposed algorithm is compared
with parametric ML estimators, specified under different scattering 
distribution assumptions, and applied onto data generated using a uniform distribution. The ML estimators
are computed under the assumption of a Gaussian, exponential, and
uniform distribution. As
expected, the best performing method is the ML estimator with the
correct distribution assumption. For the estimation of $\omega_0$, the
ML estimator with the Gaussian assumption also performs better than
our scheme. This due to the fact that both the Gaussian and uniform
shapes represent symmetric distributions. On the other hand, the
exponential assumption, which is asymmetric, yields biased results: 
asymmetry induces a bias in the estimation of $\omega_0$. The proposed
scheme is easily adapted to impose the symmetry of the
distribution by considering only the even orders. Figure~\ref{fig:
  Comparison parametric methods} also shows the RMSE obtained with
this adaptation which is, as expected, more accurate for estimating
$\omega_0$. For the estimation of the dispersion $\sigma_\omega$ and
the power $P$, the two incorrect assumptions lead to biased estimates
and are outperformed by the proposed algorithm. 

The proposed scheme is based on a approximation of the FT
$\widehat{p}$ by the first terms of its Taylor expansion at $0$. Such
an approximation is only valid if $\widehat{p}$ is sufficiently
regular in the neighborhood of 0, i.e., if the distribution $p$ is
sufficiently narrow. If the dispersion $\sigma_z$ is too large, this
approximation is no longer valid. Figure~\ref{fig: Comparison function
  of sigma} compares the performance of the schemes as a function of
$\sigma_z$. As before, it can be noted that the proposed scheme
outperforms the misspecified ML estimators over a wide range of
values. For large dispersion, when $\sigma_z$ is larger than the
resolution $\delta z \approx 14.3$ m, the performance of the
moment-based algorithm decreases as expected. For small dispersions,
the proposed approach also produces poor results, worse than the
misspecified ML estimators. There are two reasons for this: first,
when the dispersion is much smaller than the resolution, two different
distributions become hard to distinguish, and second, the proposed
scheme requires a larger number of snapshots. With more snapshots, the
proposed approach would improve its estimates, while the misspecified
ML estimators would converge to biased estimates.

\section{Concluding remarks}\label{sec: Conclusion}

This paper presents a new method for estimating the characteristics
of a spatially distributed scattering source. The proposed approach does not
assume any distribution for the scatterers, but relies on the
estimation of the central moments of their distribution. When applied
to a SAR tomography scenario, it allows to efficiently estimate the
location of the source, its power, and its dispersion without assuming
a distribution model, which would introduce estimation biases. A
second point relevant to SAR tomography applications is that the
proposed approach does not require a uniform linear array and can be
applied with irregular baselines. 
Conceptually, the moment-based estimation presented in this work is
a polynomial interpolation of the characteristic function of the
distribution. As a consequence, it is computationally efficient: the
algorithm requires only a single one-dimensional search. However, it
may suffer from the problems of overfitting or oscillations as the
resolution increases, \ie{} as $M$ increases. These problems are not
encountered in the context of SAR tomography, but they should be
further investigated in future work. Another avenue should be the
joint estimation of multiple scatterers sources, whether diffuse or
point sources.

\bibliographystyle{IEEEtran}
\bibliography{bibliography}

\end{document}